\documentclass{aa}

\usepackage{graphicx}
\usepackage{txfonts}
\usepackage{hyperref}
\defcitealias{alo18}{Paper~I}

\begin{document}

\title{A near-infrared stellar atlas of the Galactic plane from the VVVX survey}

  \author{Javier Alonso-Garc\'{i}a\inst{1,2}
  \and
  Maren Hempel\inst{3,4}
  \and
  Roberto K. Saito\inst{5}
  \and
  Dante Minniti\inst{3,6}
  \and
  Nicholas J. G. Cross\inst{7}
  \and
  Jorge Anais\inst{1}
  \and
  Jura Borissova\inst{8,2}
  \and
  M\'{a}rcio Catelan\inst{9,10,2}
  \and
  Jos\'e G. Fern\'andez-Trincado\inst{11,12}
  \and
  Elisa R. Garro\inst{13}
  \and
  Zhen Guo\inst{8,2}
  \and
  Philip W. Lucas\inst{14}
  \and
  Mar\'ia G. Navarro\inst{15}
  \and
  Casmir O. Obasi\inst{3}
  \and
  Leigh C. Smith\inst{16}
  }

  \institute{Centro de Astronom\'{i}a (CITEVA), Universidad de Antofagasta,
  Av. Angamos 601, Antofagasta, Chile\\
  \email{javier.alonso@uantof.cl}
  \and
  Millennium Institute of Astrophysics, Nuncio Monse\~nor Sotero Sanz 100,
  Of. 104, Providencia, Santiago, Chile
  \and  
  Instituto de Astrof\'isica, Departamento de F\'isica y Astronom\'ia, Facultad de Ciencias Exactas, Universidad Andr\'es Bello, Fern\'andez Concha 700, Las Condes, Santiago, Chile
  \and
  Max Planck Institute for Astronomy, K\"onigstuhl 17, 69117 Heidelberg, Germany
  \and
  Departamento de F\'{i}sica, Universidade Federal de Santa Catarina, Trindade 88040-900, Florian\'{o}polis, SC, Brazil
  \and
  Vatican Observatory, Vatican City State V-00120, Italy
  \and
  Wide-Field Astronomy Unit, Institute for Astronomy, University of Edinburgh, Royal Observatory, Blackford Hill, Edinburgh, EH9 3HJ, UK
  \and
  Instituto de F\'isica y Astronom\'ia, Universidad de Valpara\'iso, Av. Gran Breta\~na 1111, Playa Ancha, Casilla 5030, Chile
  \and
  Instituto de Astrof\'{i}sica, Facultad de F\'isica,
  Pontificia Universidad Cat\'{o}lica de Chile,
  Av. Vicu\~na Mackenna 4860, 7820436 Macul, Santiago, Chile
  \and
  Centro de Astro-Ingenier\'{i}a, Pontificia Universidad Cat\'{o}lica de Chile, Av. Vicu\~{n}a Mackenna 4860, 7820436 Macul, Santiago, Chile
  \and
  Universidad Cat\'olica del Norte, N\'ucleo UCN en Arqueolog\'ia Gal\'actica, Av. Angamos 0610, Antofagasta, Chile
  \and
  Universidad Cat\'olica del Norte, Departamento de Ingenier\'ia de Sistemas y Computaci\'on, Av. Angamos 0610, Antofagasta, Chile
  \and
  ESO - European Southern Observatory, Alonso de C\'ordova 3107, Vitacura, Santiago, Chile
  \and
  Centre for Astrophysics, University of Hertfordshire, Hatfield AL10 9AB, UK
  \and
  INAF – Osservatorio Astronomico di Roma, Via di Frascati 33, 00078, Monte Porzio Catone, Roma, Italy
  \and
  Institute of Astronomy, University of Cambridge,
  Madingley Road, Cambridge, CB3 0HA, UK
}

   \date{Received ; accepted }

  \abstract
{The VISTA Variables in the Via Lactea eXtended (VVVX) ESO public
  survey observed the Galactic plane and the outer Galactic bulge in
  the near-infrared to mitigate the effects of extinction that
  severely limit optical observations of these regions. By
  significantly expanding the area covered by the original VVV survey,
  VVVX enables a deeper and broader exploration of the most obscured and
  crowded regions of the Milky Way.}
{We aim to extend and complete our photometric catalogs of the entire
  Galactic plane region accessible from the southern hemisphere,
  focusing on the areas newly covered by the VVVX survey.}
{Building on previous work, we applied point-spread function fitting techniques to detect point sources and extract their deep
  $J$, $H$, and $K_s$ photometry across the VVVX footprint. The
  resulting catalogs were calibrated using astrometric and photometric
  reference data. Cross-matching between filters and epochs was used
  to ensure a high level of reliability and completeness.}
{We produce a deep, highly complete near-infrared catalog of more than
  700 million sources in the Galactic plane and outer Galactic
  bulge. When combined with our previous VVV atlas, the full catalog
  includes over 1.5 billion sources. The derived density maps and
  color-magnitude diagrams enable detailed studies of Galactic
  structure, extinction, and stellar populations, and highlight
  features such as the Carina arm tangency, the Sagittarius stream,
  and numerous star clusters.}
{This extended atlas provides an unprecedented view of the innermost
  regions of the Milky Way. It is now publicly available through the
  VISTA Science Archive, offering a valuable resource for the
  astronomical community to investigate the structure and evolution of
  the Galactic disk and bulge.}


   \maketitle

\section{Introduction}
\label{sec_intro}
Most of the stars in the Milky Way are located in its disk and bulge, close to the Galactic plane. However, their light at optical wavelengths is heavily obscured by clouds of gas and dust concentrated at low Galactic latitudes. The region of the Galactic plane toward the inner parts of the Galaxy is particularly affected by this obscuration. In the near-infrared, extinction is highly diminished ($A_{K_s}\sim0.1 \, A_V$), allowing observations to penetrate the gas and dust and provide a more complete and deeper view of the objects located toward the inner regions of the Milky Way.

\begin{figure*}
  \centering
  \includegraphics[scale=0.25]{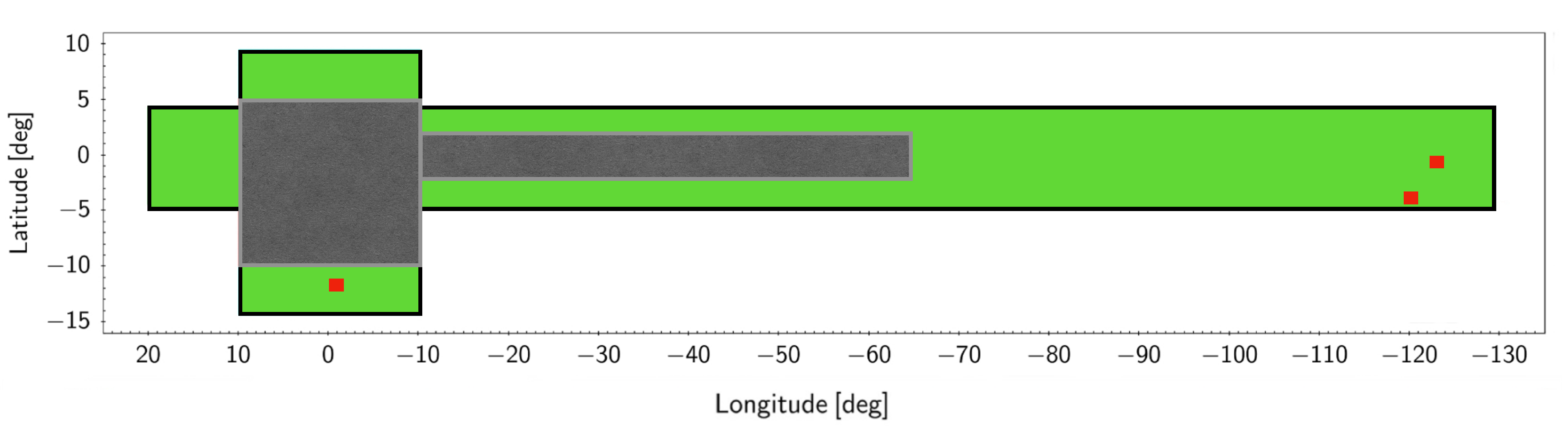}
  \caption{Footprint of the VVVX survey. The original VVV area, analyzed in \citetalias{alo18}, is shown in gray. The extended region newly covered by VVVX is shown in green. Selected fields used for the completeness tests described in Sect.~\ref{sec_psf} are marked in red.}
  \label{fig_vvvx}
\end{figure*}

The VISTA Variables in the Via Lactea (VVV) survey \citep{min10,sai12b} enabled the use of the wide-field, medium-aperture, near-infrared capabilities of the VISTA telescope and its imager to provide a deep, highly complete atlas of the Galactic bulge and an adjacent region of the southern Galactic disk, in five near-infrared filters; it contains nearly a billion sources (\citealp{alo18}, hereafter \citetalias{alo18}). The VISTA Variables in the Via Lactea eXtended (VVVX) survey \citep{min18d,sai24} subsequently expanded the original VVV footprint to cover the entire southern portion of the Galactic disk accessible from the ESO Paranal observatory in Chile (see Fig.~\ref{fig_vvvx}). This includes the fourth quadrant of the Milky Way disk, as well as a section of the first and third quadrants, complementing observations of the northern Galactic disk from the UKIDSS Galactic Plane Survey \citep[GPS;][]{luc08}. VVVX observations also extend the area of the Galactic bulge mapped by the original VVV to include its outermost regions. Although VVV and VVVX cover only approximately $4\%$ of the sky, they probe the regions of highest stellar density in the Galaxy and potentially include more than half of the Milky Way's stars \citep{min18d}.

Point-spread function (PSF) fitting photometry is specifically designed to extract the photometry of point sources located in highly crowded regions \citep{hea99}, such as those covered by the VVV and VVVX surveys. In \citetalias{alo18} we applied this technique to analyze and produce a near-infrared atlas of the sources within the VVV footprint, demonstrating that it reaches a greater depth and achieves a higher completeness than aperture photometry. In the present paper, we extend this analysis to the entire VVVX footprint, providing a $J$, $H$, and $K_s$ near-infrared atlas of the full portion of the Galactic plane observable from the southern hemisphere.    

\section {Observations}
\label{sec_obs}
The VVV and the VVVX observations were carried out with the 4.1m VISTA telescope located in ESO's Cerro Paranal Observatory in Chile. As mentioned in \citetalias{alo18}, the VIRCAM camera on the telescope provided wide-field, sparsely sampled images covering a $1.5^\circ \times 1^\circ$ area of the sky, with a resolution of $0.34^{\prime\prime}$ per pixel. The significant gaps between the 16 detectors in the camera result in a noncontiguous coverage of the field of view. The observing strategy for the VVVX observations \citep{sai24} was similar to that employed in the VVV survey: two consecutive, slightly jittered exposures are taken and combined into so-called stacked pawprints to correct for detector cosmetic effects. Then, to achieve contiguous coverage of the observed field, a mosaic of six stacked pawprints is used to fill the gaps between the detectors, resulting in a so-called tile. Unlike the VVV strategy, however, the VVVX survey provides observations only in the $J$, $H$, and $K_s$ near-infrared filters.

The VVVX survey covers almost 1700 square degrees in the Galactic bulge and disk \citep{sai24}, approximately three times the area covered by the VVV survey (see Fig.~\ref{fig_vvvx}). 
While the original VVV footprint included the Galactic plane between  $-65\fdg3 \le l\le10\fdg0$ and $-2\fdg25 \le b \le 2\fdg25$, the VVVX footprint extends this region in both longitude and latitude, reaching $-130\fdg6 \le l\le20\fdg25$ and $-4\fdg4 \le b \le 4\fdg4$. Additionally, the original VVV footprint of the Galactic bulge ($-10\fdg0 \le l \le +10\fdg4$ and $-10\fdg3 \le b \le +5\fdg1$) is expanded in VVVX to include its outermost regions ($-10\fdg0 \le l \le +10\fdg6$ and $-14\fdg6 \le b \le +9\fdg4$). In total, we surveyed 1028 VISTA fields: the 348 fields in the original VVV footprint (gray area in Fig.~\ref{fig_vvvx}), plus 568 new fields in the Galactic plane and 112 new fields in the outer Galactic bulge (green region in Fig.~\ref{fig_vvvx}). The effective exposure times for the stacked pawprints were the same for all VVVX fields: 60 seconds in $J$, 24 seconds in $H$, and 8 seconds in $K_s$. However, with a few exceptions, the fields covered in the original VVV footprint were imaged in fewer than ten epochs in $K_s$, primarily to improve the time coverage for their variable sources and to enhance proper-motion precision by extending the time baseline; only the very low-latitude VVV bulge fields ($-2\fdg4 < b < 1\fdg5$) were also reobserved in J and H (see Table A.1 in \citealt{sai24}) to improve the color-magnitude depth in these highly reddened regions. The VVVX fields outside the VVV footprint were observed in the $J$, $H$, and $K_s$ filters -- typically with two epochs in $J$, one epoch in $H$, and between 23 and 50 epochs in $K_s$, where the variability campaign was conducted.

\section{PSF photometry and catalogs}
\label{sec_psf}
\begin{figure}
  \centering
  \includegraphics[scale=1.0]{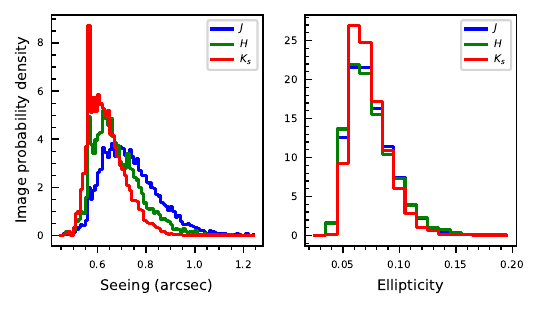}
  \caption{Seeing (\textit{left}) and ellipticity (\textit{right}) of the sources in the VVVX images used to construct our PSF near-infrared atlas. $J$-band images are shown in blue, $H$-band in green, and $K_s$-band in red.}
   \label{fig_quality}
\end{figure}

In \citetalias{alo18} we present a catalog with the photometry of nearly a billion sources within the original VVV footprint. In the present work, we aim to find and extract PSF photometry of point sources in the VVVX regions not previously surveyed by VVV (shown in green in Fig.~\ref{fig_vvvx}). To construct our near-infrared atlas, we used the two available epochs per field in the $J$ filter, the single available epoch in $H$, and selected two $K_s$ epochs per field with good seeing conditions (i.e., better than $1^{\prime\prime}$). As shown in Fig. \ref{fig_quality}, the quality of the VVVX images used in our analysis was excellent, with a mean seeing of $\sim0.6-0.7^{\prime\prime}$ and typical ellipticities below 0.1. The higher average quality of the $K_s$-band images compared to the other filters is primarily due to the larger number of available epochs, as the variability campaign was conducted in this filter.

The initial steps of our analysis followed those described in \citetalias{alo18}. We first ran DoPHOT, the software we used to detect the sources and extract their PSF photometry \citep{sch93,alo12}, on each of the detectors in the stacked pawprints. As mentioned in \citetalias{alo18}, we used the stacked pawprints because they provide a higher signal-to-noise ratio than single pawprints, correct many of the cosmetic defects present in individual detectors, and avoid the PSF variations seen in VISTA tiles.

The reduced VVVX stacked pawprint images and the corresponding catalogs used for astrometric and photometric calibration were provided by the Cambridge Astronomy Survey Unit (CASU). To convert the instrumental pixel positions of the DoPHOT-detected sources into equatorial coordinates, we used WCSTools \citep{mink19,mink06} and the astrometric solutions provided by CASU in the stacked pawprints. These solutions incorporate the VIRCAM radial distortion model \citep{sai12b}, and are tied to the 2MASS point-source catalog \citep{gonfer18}, enabling a very small rms of $\sim70$ mas in the resulting world coordinate system \citep{sai12b}. For photometric calibration into the VISTA magnitude system \citep{gonfer18}, we selected the brightest non-saturated stars detected by DoPHOT in each detector, filter, and epoch, matched them with the calibrated CASU catalogs, computed the average zero point, and applied it to all sources detected in the corresponding detector.

Using STILTS \citep{tay06}, we then combined the photometry from the mosaic of six stacked pawprints per filter and epoch, retaining all detected sources and applying a weighted average (based on photometric errors) for those sources detected more than once. Subsequently, we used STILTS again to combine the photometry from the two $J$-band epochs, keeping only sources detected in both epochs. We computed the weighted-average magnitude for these sources and retained only those for which the averaged magnitude was within $3{\sigma}$ of the individual epoch magnitudes. The same procedure was applied to the two selected $K_s$-band epochs. No epoch combination was performed for the $H$ filter due to the availability of only a single observation per field in this band.

Next, we cross-matched the $J$, $H$, and $K_s$ photometry using STILTS, retaining only sources detected in at least two of these bands. This multi-epoch, multiband filtering strategy significantly reduced the number of spurious detections, especially those occurring near saturated stars. To further eliminate residual spurious sources, we performed a smooth, nonparametric fit to the photometric errors as a function of magnitude for each filter, and excluded any sources deviating by more than $10{\sigma}$ from the fitted trend. This process yielded a near-infrared source catalog for each of the 680 VVVX fields located outside the original VVV footprint.

\begin{figure}
  \centering
  \includegraphics[scale=1.0]{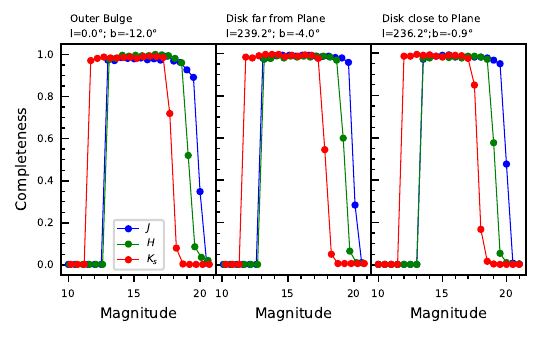}
  \caption{Completeness profiles for representative fields in the VVVX outermost bulge (\textit{left}), the VVVX disk outside the Galactic plane (\textit{middle}), and the VVVX disk near the Galactic plane (\textit{right}). $J$-band results are shown in blue, $H$-band in green, and $K_s$-band in red.}
  \label{fig_comp}
\end{figure}
\begin{figure}
  \centering
  \includegraphics[scale=1.0]{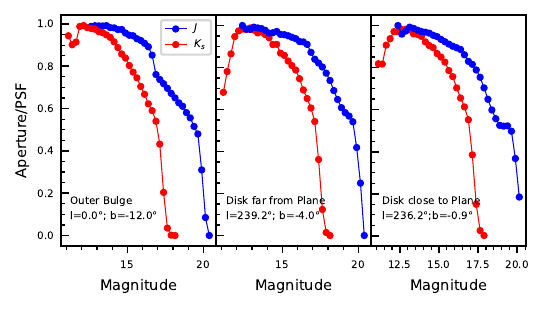}
  \caption{Ratio of the number of detected sources as a function of magnitude between our PSF photometry and CASU aperture photometry catalogs, for representative fields in the VVVX outermost bulge (\textit{left}), the VVVX disk outside the Galactic plane (\textit{middle}), and the VVVX disk close to the Galactic plane (\textit{right}). $J$-band results are shown in blue, and $K_s$-band in red.}
  \label{fig_comp2}
\end{figure}

All VVVX field catalogs were integrated into a single table in the VISTA Science Archive (VSA), named {\it vvvxPsfDophotJHKsSource}, which constitutes the final VVVX atlas. Only the combined, field-level catalog is provided in the archive; individual-detection catalogs are not included. Each entry in \textit{vvvxPsfDophotJHKsSource} corresponds to a unique point source and contains its equatorial coordinates, computed as weighted averages from the multi-epoch cross-matching, together with the $J$, $H$, and $K_s$ magnitudes and their associated uncertainties, also derived as weighted averages. Since adjacent VVVX fields overlap by a few arcminutes, the table additionally includes the {\it priOrSec} attribute, as in \citetalias{alo18}, which enables users to construct seamless, non-duplicated catalogs across overlapping fields.

A concise description of all columns and their data types is available in the VSA Schema Browser\footnote{\url{http://vsa.roe.ac.uk/www/vsa_browser.html}}  in VVVX Data Release 1. Users can, for instance, select all primary (non-duplicate) sources in a given field, retrieve photometry within a specified magnitude range, or extract color–magnitude diagrams (CMDs) through simple SQL queries, with several example queries provided in the VSA documentation.

Our resulting atlas contains near-infrared photometry for more than 707 million individual sources detected in the VVVX region not previously imaged in the VVV -- 550 million in the Galactic plane, and over 157 million in the outer Galactic bulge.

Figures~\ref{fig_densitybulge}~and~\ref{fig_densitydisk} present the source density maps for the outer bulge and Galactic plane, respectively. As expected, regions near the Galactic plane and bulge exhibit the highest densities of detected sources. However, the distribution is not uniform and reveals a highly patchy structure, shaped by the variable presence of interstellar gas and dust. While near-infrared observations significantly reduce the extinction effects on background point sources, they do not eliminate them entirely. These differential extinction patterns, however, allow us to trace the morphology of gas and dust clouds in great detail (see the upper panel of Fig.~\ref{fig_densitydisk}).

Moving away from the Galactic plane, the source densities decrease markedly (see Figs.~\ref{fig_densitybulge}~and~\ref{fig_densitydisk}). A similar trend is observed with increasing distance from the Galactic bulge, where point-source densities gradually decline (see the central panel of Fig.~\ref{fig_densitydisk}). We also identify a sharp change in stellar density at very low latitudes and at longitudes around $l\sim285^\circ$ (or equivalently $l\sim-75^\circ$; see the bottom panel of Fig.~\ref{fig_densitydisk}). We are currently exploring its possible association with the Carina arm tangency, previously observed using other tracers but not mapped using stellar populations \citep{hou15,rus23}.

The presence of numerous star clusters along the Galactic plane, as well as several densely populated globular clusters in the outer bulge, is revealed as localized peaks in source density in  Figs.~\ref{fig_densitybulge}~and~\ref{fig_densitydisk}. Finally, in the lower panel of Fig.~\ref{fig_densitybulge}, we note the presence of the Sagittarius stream as a band of higher source density at mean Galactic longitude $l\sim5\fdg5$, surrounding the globular cluster M~54.

Because the VVVX survey covers diverse Galactic regions, we selected three representative fields from the newly surveyed area (highlighted in red in Fig.~\ref{fig_vvvx}: one located in the extended Galactic bulge, centered at $l=-0\fdg4$, $b=-11\fdg9$; a second in the Galactic plane but far from the bulge, at $l=-124\fdg1$, $b=-0\fdg6$; and a third also far from the bulge, in the extended Galactic disk, at $l=-121\fdg2$, $b=-3\fdg8$). To assess photometric completeness, we conducted artificial star tests by injecting batches of 5000 synthetic stars into the images, using 0.5 magnitude bins spanning the full detection range in each filter. The results of these tests are shown in Fig.~\ref{fig_comp}. Across the whole dynamic range -- approximately 6.5 mag in $J$ and 5.5 mag in $H$ and $K_s$ -- we recovered highly complete samples of the injected stars. However, completeness decreases rapidly at the faint end, dropping from over $90\%$ to $0\%$ within just one magnitude. We find that the completeness trends are similar across all three examined fields, despite their widely varying Galactic locations, and that they are higher than those in the original VVV footprint (see Fig.~4 in \citetalias{alo18}). This suggests that the newly surveyed VVVX regions are not as extreme in terms of stellar crowding and extinction as those observed in VVV. Nevertheless, as shown in Fig.~\ref{fig_comp2}, our PSF photometry recovers significantly more point sources than the aperture photometry available in the CASU catalogs.

The main caveats of the catalogs produced for the VVVX are the same as those discussed in \citetalias{alo18} for the VVV catalogs. Most notably, variable sources whose magnitudes differ by more than $3{\sigma}$ between the two available epochs in a given filter are excluded, as this was the threshold adopted for retaining a source in the final catalog. In addition, the catalogs will lack stars with very high proper motions, as these may fall outside the cross-matching tolerance of $0.34^{\prime\prime}$ (approximately one pixel) when cross-correlating the photometry between the three VVVX filters.   

\section{The VVVX CMDs}
\label{sec_cmd}
\begin{figure}
  \centering
  \includegraphics[scale=0.33]{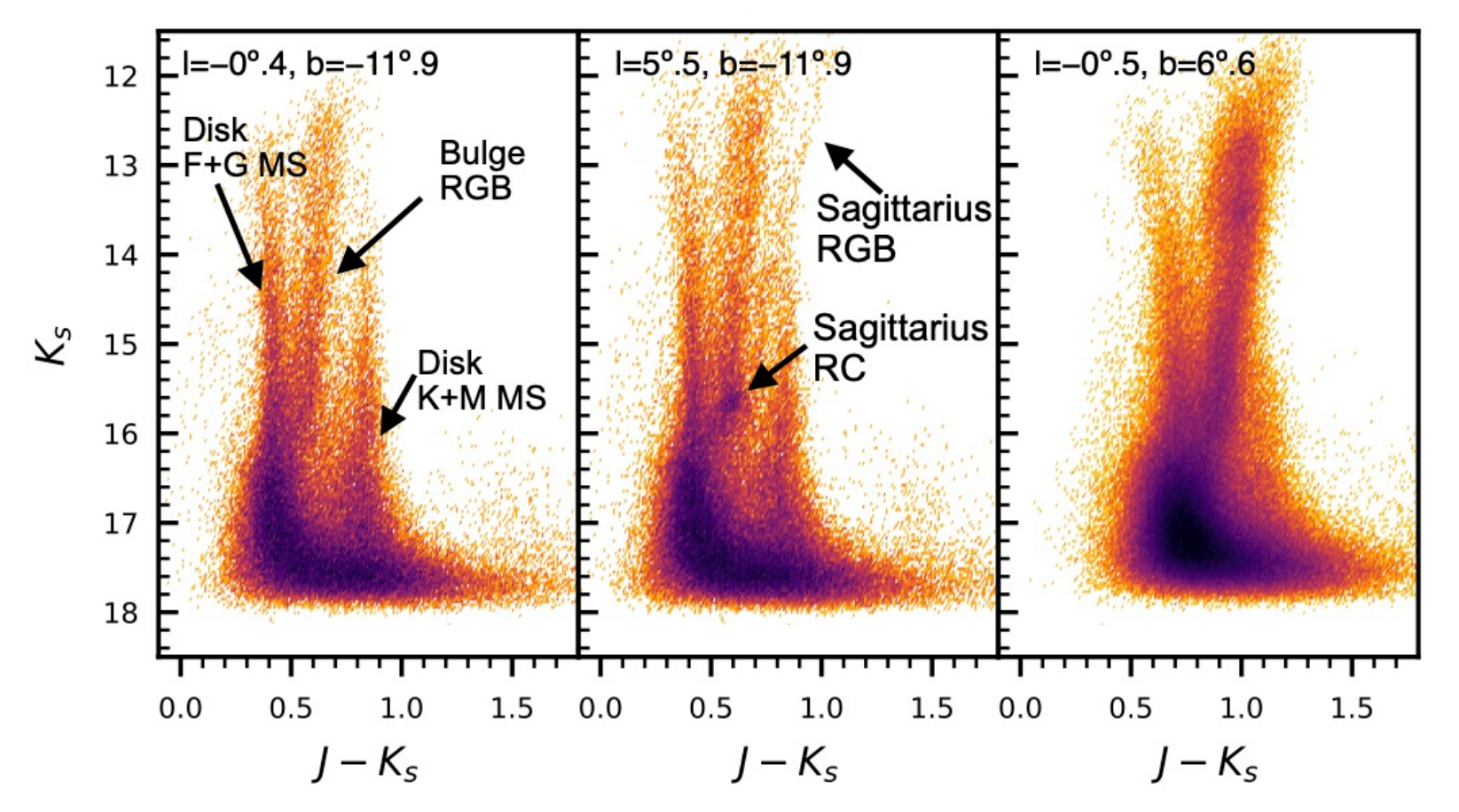}
  \caption{Near-infrared CMDs of selected regions in the outer Galactic bulge within the VVVX footprint. Darker colors indicate higher source densities. Each field corresponds to a circular region within $r<15^\prime$ of the coordinates shown above the panels.}
  \label{fig_cmd_bulge}
\end{figure}

\begin{figure}
  \centering
  \includegraphics[scale=0.33]{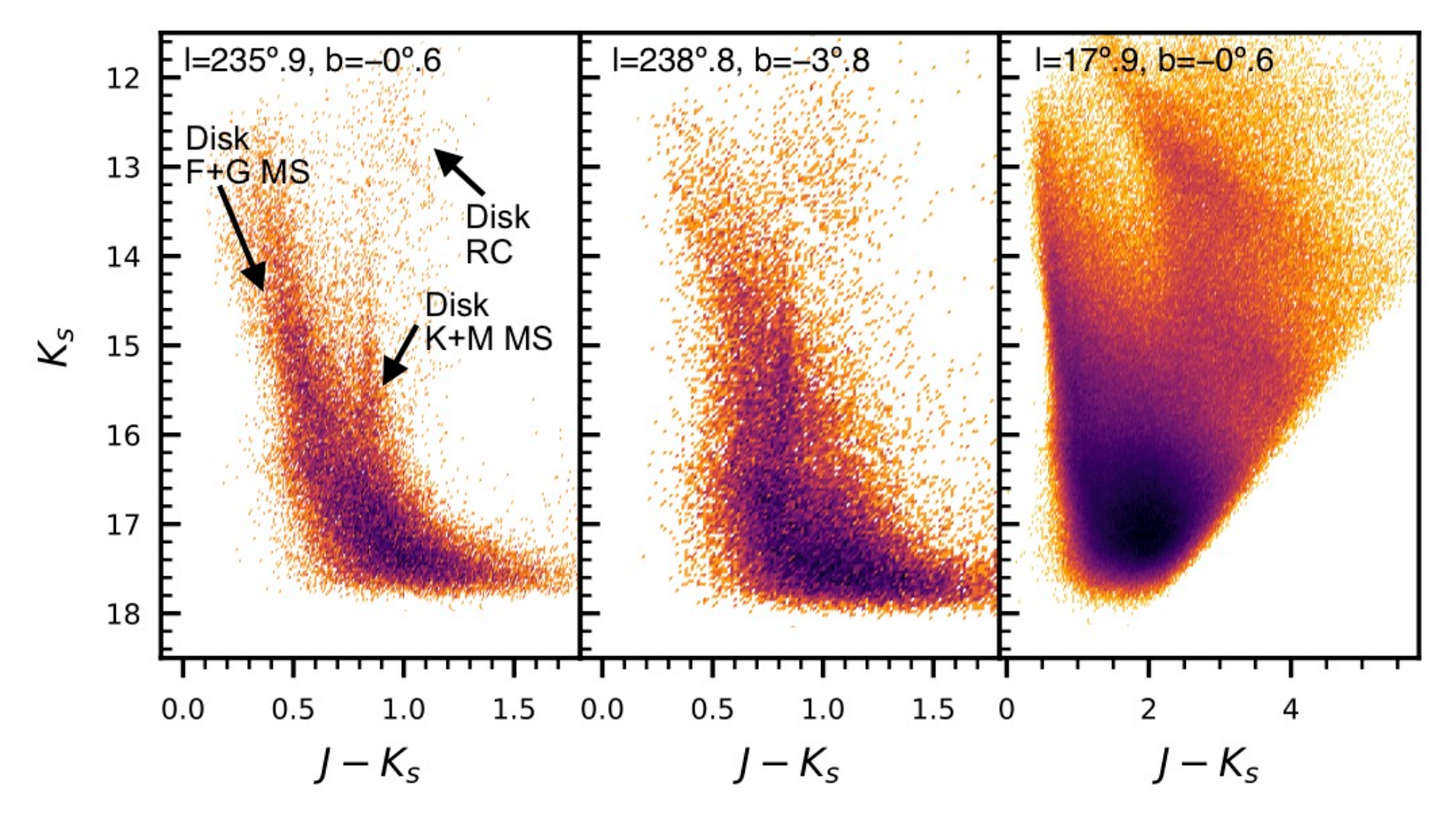}
  \caption{Near-infrared CMDs of selected regions in the Galactic disk within the VVVX footprint. Darker colors indicate higher source densities. Each field corresponds to a circular region within $r<15^\prime$ of the coordinates shown above the panels.}
  \label{fig_cmd_disk}
\end{figure}

Given the range of Galactic components covered by VVVX, we find it more appropriate in this section to focus on the information that can be extracted from the CMDs of selected regions within the surveyed area. To this end, we plotted CMDs for regions within a radius of $15^\prime$ centered on representative positions: three located in the VVVX outer bulge (see Fig.~\ref{fig_cmd_bulge}) and other three in the VVVX Galactic disk (see Fig.~\ref{fig_cmd_disk}). We also present CMDs for stars identified in the inner regions of two reference stellar clusters (see Fig.~\ref{fig_cmd_clusters}). In the following, we discuss the main evolutionary sequences and features identified in these CMDs, along with the stellar populations responsible for them.

\subsection{The CMDs of the outer Galactic bulge fields in VVVX}
\label{sec_cmdBulge}
As mentioned in Sect.~\ref{sec_obs}, in addition to the Galactic bulge regions covered by the original VVV footprint, VVVX extends coverage to two additional areas in the outer bulge (see Figs.~\ref{fig_vvvx}~and~\ref{fig_densitybulge}). Figure~\ref{fig_cmd_bulge} shows the CMDs for three representative fields located in these regions. In these CMDs, red giant branch (RGB) stars belonging to the bulge dominate at magnitudes $K_s<16$ and colors $(J-K_s)>0.5$. The CMDs also reveal the presence of foreground main sequence (MS) stars from the Galactic disk: there is an almost vertical sequence at $(J-K_s)\sim0.9$, corresponding to relatively nearby disk late K- and M-type MS stars; additionally, a second sequence, comprising bluer F- and G-type MS stars from the disk, overlaps with the subgiant branch and upper MS of the bulge at the faintest magnitudes ($K_s>16$).

The northern region of the outer bulge sampled by VVVX (right panel in Fig.~\ref{fig_cmd_bulge}) lies closer to the Galactic plane than the southern region (left and central panels in Fig.~\ref{fig_cmd_bulge}). As a result, the number of stars is significantly higher -- by a factor of four in the CMDs we show in this figure (see also the density maps in Fig.~\ref{fig_densitybulge}). Due to the difference in Galactic latitude, both absolute and differential reddening are also more significant in the northern region than in the southern area. This is evident in the CMDs in the right panel of Fig.~\ref{fig_cmd_bulge}, where the stellar sequences appear both redder and broader. However, since both regions lie at relatively high latitudes ($|b|<5^\circ$), the extinction is not as severe as in areas located closer to the Galactic plane, as illustrated by the CMDs of the inner Galactic bulge in the right panels of Fig.~6 in \citetalias{alo18}.     

Finally, as detailed in Sect.~\ref{sec_psf}, there is a region in the southern outer bulge where the Sagittarius stream can be readily identified in our data. The CMD for a field selected in this area is shown in the central panel of Fig.~\ref{fig_cmd_bulge}. In this CMD, the RGB of the Sagittarius stream appears as a distinct sequence, slightly redder than the RGB of the Galactic bulge. An overdensity of stars at $K_s\sim15.7$ and $(J-K_s)\sim0.6$, corresponding to the Sagittarius red clump, is also evident. These features are not present in fields at similar latitudes but located far from the stream (see the left panel of Fig.~\ref{fig_cmd_bulge}). We are currently working to isolate the Sagittarius population from the Galactic stars using the VVVX catalogs, with the goal of characterizing the Sagittarius stream at low Galactic latitudes (Anais et al., in prep).   

\subsection{The CMDs of the outer Galactic disk in VVVX}
\label{sec_cmdDisk}
As detailed in Sect.~\ref{sec_obs}, the VVVX survey significantly extends the coverage of the Galactic disk beyond that of the original VVV. Figure~\ref{fig_cmd_disk} presents the CMDs for three representative fields in the Galactic disk. The left and central panels correspond to regions at longitudes beyond the Carina arm tangency \citep[$l = 282^{\circ}$ (–78$^{\circ}$);][]{rus23}, which results in a significantly lower density of detected stars in the CMDs (see the lower panel of Fig.~\ref{fig_densitydisk}). Nevertheless, the nearly vertical sequence of nearby K- and M-type MS stars from the disk is clearly visible at $(J-K_s)\sim0.9$. At fainter magnitudes, this sequences blends with the bluer sequence of the more distant F- and G-type disk MS stars. A redder sequence at brighter magnitudes ($K_s<14$, $(J-K_s)>0.9$) corresponding to disk red clump stars is also discernible in these two CMDs. Although these two fields are relatively close in projection, the field farther from the plane exhibits noticeably higher extinction, as evidenced by the broadening and reddening of both the distant F- and G-type MS stars and the red clump sequence. The CMD in the right panel, corresponding to disk stars in the first quadrant, shows a significantly higher stellar density. However, the evolutionary sequences are less well defined due to the stronger differential extinction in this region. 

\begin{figure}
  \centering
  \includegraphics[scale=0.33]{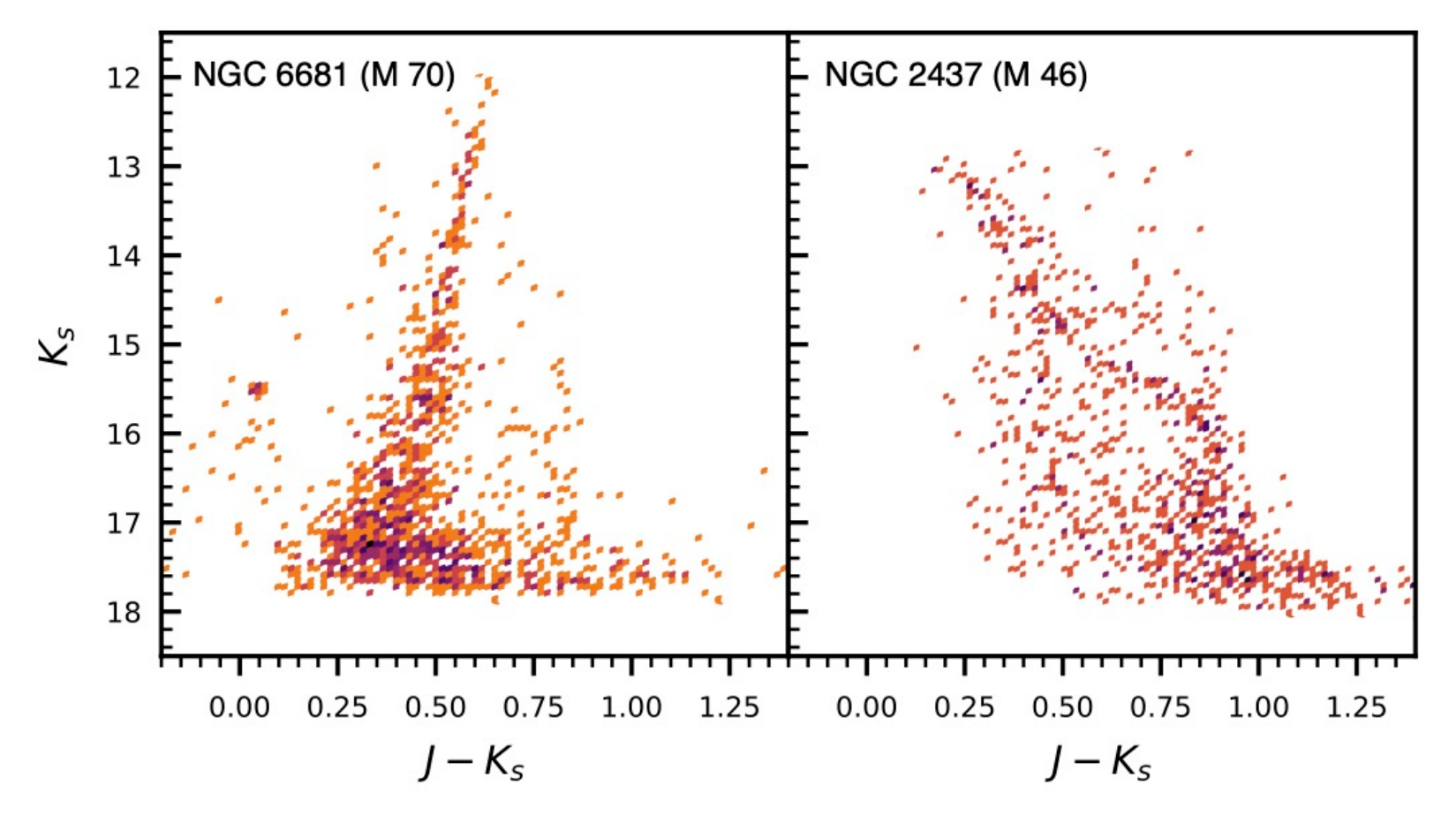}
  \caption{Near-infrared CMDs of the inner regions of two selected Galactic stellar clusters. \textit{Left}: Globular cluster NGC~6681 (M~70), within $r<2.5'$. \textit{Right}: Open cluster NGC~2437 (M~46), within $r<5'$.}
  \label{fig_cmd_clusters}
\end{figure}

\subsection{The CMDs of stellar clusters in VVVX}
\label{sec_cmdClusters}
As stellar clusters are highly crowded objects, their study toward the Galactic plane can greatly benefit from the use of our PSF near-infrared photometric atlas. As an illustration, Fig.~\ref{fig_cmd_clusters} shows stars identified in the inner regions of the globular cluster M~70 and the open cluster M~46. The most evolved sequences (RGB and horizontal branch) are visible in M~70, while for M~46, the lower MS extending down to the substellar sequence can be observed.

Based on detailed analysis of our near-infrared atlas, several studies from our collaboration have identified new embedded and open cluster candidates \citep{bor18,bor19,bor20}, as well as globular cluster candidates \citep{min20,gar20,gar22b,gar22c,gar24,oba21,sar24}. While many of these clusters are compact and faint, the study of some larger and more massive globular clusters has also benefited from our database \citep{min18c,bar19}. Interestingly, some of the newly identified globular clusters candidates appear to be associated with the Sagittarius dwarf galaxy \citep{min21b,min21a,gar21b,gar22a} and with even more distant galaxies \citep{oba23}.

\section{Conclusions}
\label{sec_conclusions}
The VVVX survey extends the footprint of the original VVV survey to cover the entire Galactic plane observable from the southern hemisphere up to $|b|<4\fdg4$, while also expanding the coverage around the Galactic bulge to include its outermost regions. The area surveyed is approximately three times larger than that of the original VVV footprint, enabling a deep view into these highly extincted and crowded regions of our Galaxy.

In this paper we present a near-infrared atlas of the new regions covered by VVVX, which, when combined with the atlas published in \citetalias{alo18} for the original VVV footprint, provides a catalog of more than 1.5 billion point sources. This combined dataset offers an unprecedented view of the most obscured and densely populated regions of the Galactic bulge and disk.

As expected, the source density maps reveal higher densities toward the Galactic plane and bulge; however, the distributions are not uniform, but rather highly patchy due to the effects of extinction. We also present CMDs of selected reference regions, where we identify the different stellar populations found at these low Galactic latitudes.

This atlas is already being used within our collaboration for a variety of scientific investigations, including tracing the spiral arm tangencies, identifying low extinction windows, mapping the stellar populations of the Sagittarius stream at low latitudes, and discovering new star clusters in the Galactic bulge and disk. It will also serve as a valuable complement to upcoming photometric surveys of the Galactic plane, such as those planned with the \textit{Vera C. Rubin} Observatory and the \textit{Nancy Grace Roman} Space Telescope. Furthermore, it will provide promising targets for near-infrared spectroscopic surveys, including the recently approved KMOS VVVX-GalCen Spectroscopic Survey. The atlas is now publicly available through the VSA for use by the broader astronomical community.

\begin{acknowledgements}
We gratefully acknowledge the use of data from the ESO Public Survey
program ID 198.B2004, taken with the VISTA telescope, and data
products from the Cambridge Astronomical Survey Unit.  J.A.-G., J.B.,
M.C., and Z.G. are supported by ANID's Millennium Science Initiative
through grants ICN12\textunderscore009 and AIM23-0001, awarded to the
Millennium Institute of Astrophysics (MAS). J.A.-G. also acknowledges
support from ANID's FONDECYT Regular grant No. 1201490, and support
from DGI-UAntof and Mineduc-UA Cod. 2355.  R.K.S. acknowledges support
from CNPq/Brazil through projects 308298/2022-5 and
421034/2023-8. D.M. acknowledges support by ANID Fondecyt Regular
grant No. 1220724, and by the BASAL Center for Astrophysics and
Associated Technologies (CATA) through ANID grants ACE210002 and AFB
210003. J.B. thanks the support from the ANIDs FONDECYT Regular grant
No. 1240249. Support for M.C. is also provided by ANID’s FONDECYT
Regular grant \#1231637 and Basal project FB210003. J.G.F-T gratefully
acknowledges the grants support provided by ANID's FONDECYT
Iniciaci\'on No. 11220340, ANID Fondecyt Postdoc No. 3230001
(Sponsoring researcher), from the Joint Committee ESO-Government of
Chile under the agreement 2021 ORP 023/2021 and 2023 ORP
062/2023. Z.G. also acknowledges the support from the China-Chile
Joint Research Fund (CCJRF No.2301) and the Chinese Academy of
Sciences South America Center for Astronomy (CASSACA) Key Research
Project E52H540301. P.W.L. acknowledges support by STFC grant
ST/Y000846/1. C.O.O. acknowledge the funding by the Postdoctoral
Talent Attraction Competition for Research Centers and Institutes of
the Universidad Andr\'es Bello(UNAB) 2025, project No. DI-07-25/ATP.
\end{acknowledgements}

\bibliographystyle{aa} 
\bibliography{mybibtex}

\begin{appendix}
  \begin{figure*}
    \section{Point-source density maps of the VVVX surveyed regions}
    \centering
    \includegraphics{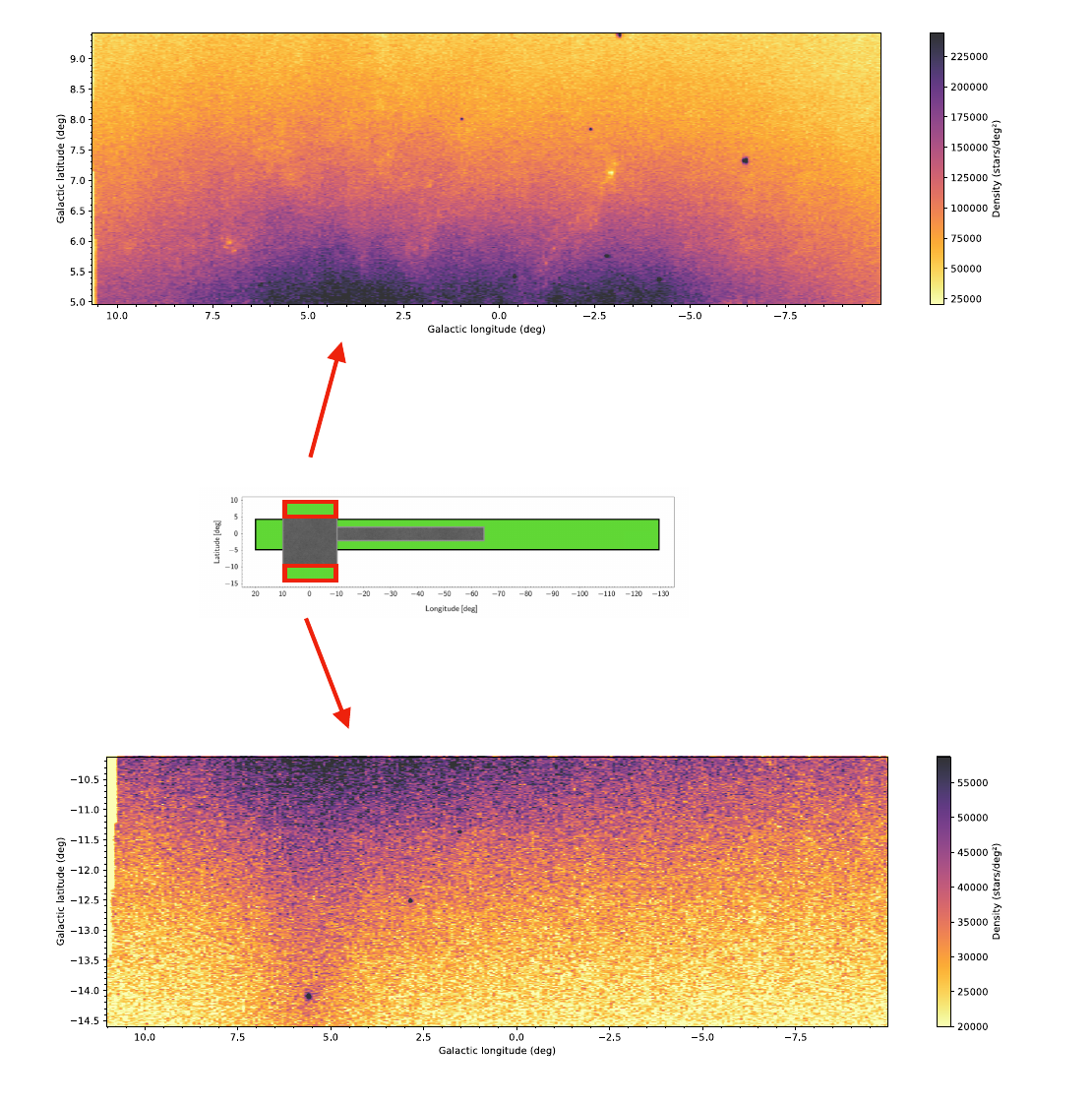}
    \caption{Point-source density maps of the VVVX surveyed regions in the outer bulge, showing sources detected down to $K_s=16.0$. Higher densities are shown with cooler colors, as indicated by the color bars on the right. Density increases toward the Galactic plane; note the different color bar ranges for the two panels. Darker points mark highly crowded stellar clusters. In the lower panel, the Sagittarius stream appears as a region of enhanced density near Galactic longitude $l\sim5\fdg5$.}
    \label{fig_densitybulge}
  \end{figure*}

\begin{figure*}
  \centering
  \includegraphics[scale=0.85]{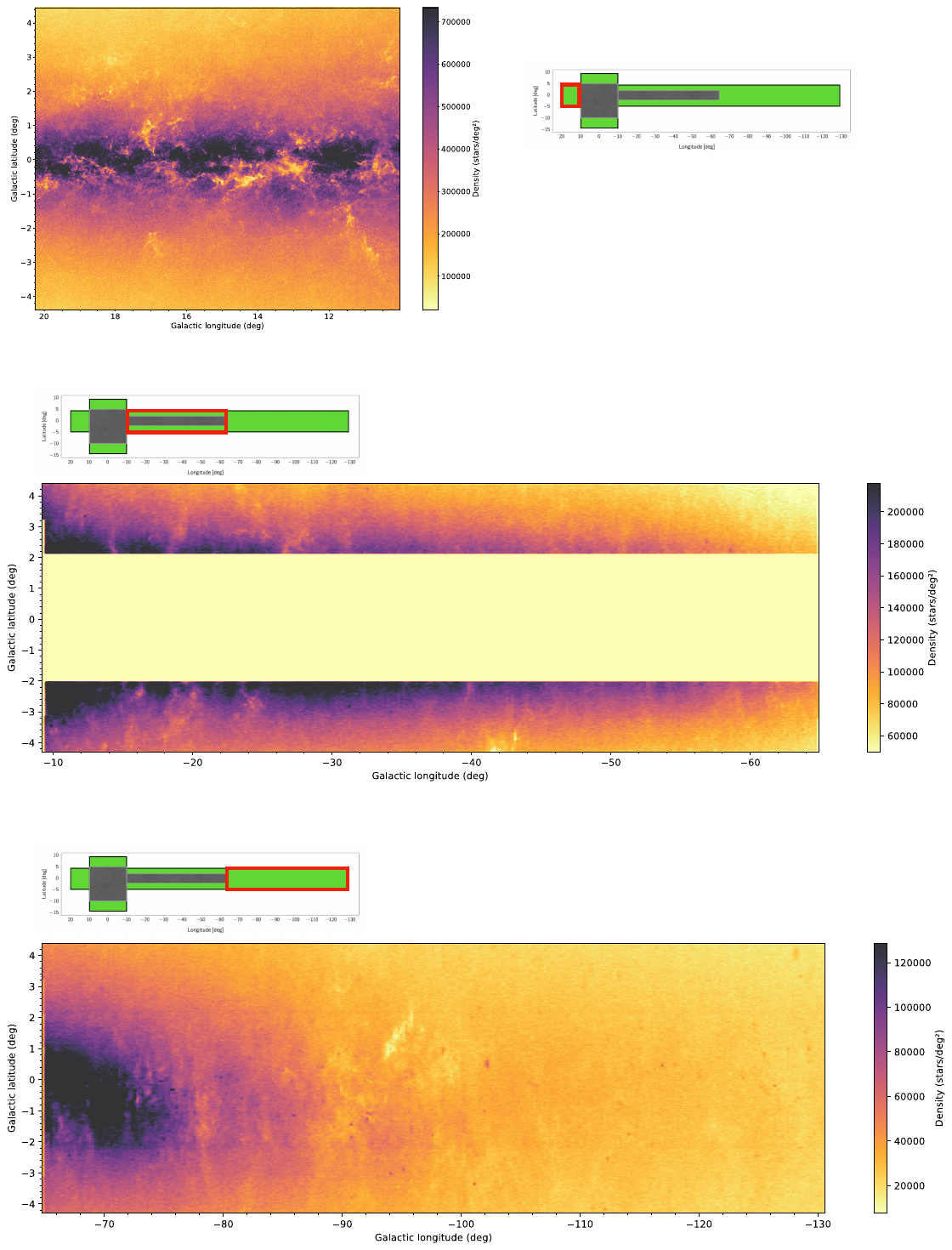}
  \caption{Same as Fig.~\ref{fig_densitybulge} but for the VVVX surveyed regions in the Galactic plane. Regions closer to the Galactic bulge, shown in the upper and middle panels, exhibit higher source densities. The abrupt drop in source density in the lower panel aligns with the position of the Carina arm tangency.}
  \label{fig_densitydisk}
\end{figure*}

\end{appendix}

\end{document}